# Language Models for Code Completion: A Practical Evaluation

Maliheh Izadi
m.izadi@tudelft.nl
Delft University of Technology
Delft, Netherlands

Jonathan Katzy
j.b.katzy@tudelft.nl
Delft University of Technology
Delft, Netherlands

Tim van Dam
t.o.vandam@student.tudelft.nl
Delft University of Technology
Delft, Netherlands

Marc Otten
m.j.c.otten@student.tudelft.nl
Delft University of Technology
Delft, Netherlands

Razvan Mihai Popescu
r.popescu-3@student.tudelft.nl
Delft University of Technology
Delft, Netherlands

Arie van Deursen
arie.vandeursen@tudelft.nl
Delft University of Technology
Delft, Netherlands

## ABSTRACT

Transformer-based language models for automatic code completion have shown great promise so far, yet the evaluation of these models rarely uses real data. This study provides both quantitative and qualitative assessments of three public code language models when completing real-world code. We first developed an open-source IDE extension, *Code4Me*, for the online evaluation of the models. We collected real auto-completion usage data for over a year from more than 1200 users, resulting in over 600K *valid* completions. These models were then evaluated using six standard metrics across twelve programming languages. Next, we conducted a qualitative study of 1690 real-world completion requests to identify the reasons behind the poor model performance. A comparative analysis of the models' performance in online and offline settings was also performed, using benchmark synthetic datasets and two masking strategies.

Our findings suggest that while developers utilize code completion across various languages, the best results are achieved for mainstream languages such as Python and Java. InCoder outperformed the other models across all programming languages, highlighting the significance of training data and objectives. Our study also revealed that offline evaluations do not accurately reflect real-world scenarios. Upon qualitative analysis of the models' predictions, we found that 66.3% of failures were due to models' limitations, 24.4% occurred due to inappropriate model usage in a development context, and 9.3% were valid requests that developers overwrote. Given these findings, we propose several strategies to overcome the current limitations. These include refining training objectives, improving resilience to typographical errors, adopting hybrid approaches, and enhancing implementations and usability.

## KEYWORDS
Automatic Code Completion, Transformers, Language Models, IDE, Evaluation, Open Source, InCoder, UniXcoder, CodeGPT

## 1 INTRODUCTION

The rising popularity of transformer-based Language Models (LMs) has significantly impacted development practices, with more developers increasingly adopting these models for code completion. Unlike traditional native auto-completions, LMs work well in predicting entire lines of code [1–12]. Despite their advantages, LMs for code are not without flaws. Many LMs proposed in research are complex and accurate on synthetic data [13], but only a handful ever make it to integration in products, highlighting a gap between theoretical advancement and practical application [7, 14–16]. A few studies have analyzed these models in online settings [17, 18], revealing that their performance may decrease in such environments. However, these explorations often involve proprietary models and code in specialized scenarios, and they fall short of providing a detailed analysis of the errors. Thus, a more thorough, comparative, and open evaluation of LMs for code, especially in real-world contexts, is critical for improving their effectiveness.

The goal of this study is to systematically evaluate three public completion models, namely InCoder [2], UniXcoder [3], and CodeGPT [19] in a real-world coding setup. We released *Code4Me*, an open-source IDE extension, to conduct *online* evaluations. Over the course of a year, we gathered auto-completion usage data from more than 1200 users, leading to about 2M completions. We examine the three models across twelve programming languages commonly used in the literature [13, 20, 21] using six standard metrics. Additionally, we carry out a *qualitative* investigation using an open coding process on 1690 real completion requests to identify the causes of poor model performance. Lastly, we conduct a comparison of the models' performance in both online and *offline* settings, employing benchmark synthetic datasets [13] and two distinct masking strategies used in literature [1, 21].

Analysis of the models' online use reveals their application across diverse programming languages. However, performance varies by language. Models excel in common languages like Python and Java but face challenges with less-represented languages like Rust and Scala. Furthermore, InCoder exceeds its competitors across all languages, highlighting the impact of training data and more aligned training objectives as it was trained for code infilling. Our analysis identified 18 types of rejections by developers. Often, the models' design limited their accuracy or developers expected predictions beyond the models' scope. We also noted a tendency of developers to reject correct predictions. Lastly, we found a mismatch between model performance on synthetic and real-world data. This is likely due to the different contexts in generating predictions for incomplete code, versus the complete code used in training.

Based on our findings, we recommend that future research focus on (1) including a broader range of programming languages in both the training and evaluation phases to achieve more consistent performance, (2) refining model training objectives to more accurately reflect actual developer usage, and (3) improving assistant implementations and usability as well as reducing unnecessary invocations. Our contributions are as follows.

- Code4Me, an open-source IDE extension that provides auto-completions using three code LMs, with a substantial user base (+1200) and nearly 2M completions,
- Quantitative online and offline evaluation of the models across twelve programming languages (+600K valid completions),
- An analysis of models' limitations using 1690 completions resulting in a taxonomy of 18 causes of poor performance,
- Public source code, dataset for offline evaluation, and open coding data for the qualitative analysis [22].

## 2 BACKGROUND AND RELATED WORK

### 2.1 Transformer-based Language Models

Transformer-based LMs require substantial datasets for initial pre-training and potential fine-tuning for specific tasks. Common architectures are encoder-only transformers such as BERT [23, 24], decoder-only transformers such as GPT [25–27], or encoder-decoder models such as T5 [28]. Popular software engineering downstream tasks with transformer-based models are code interpreting and representation [4, 5, 9, 29, 30], source code and software documentation [2, 4, 5, 31, 32], and code generation [1, 2, 4–7, 10, 13, 33].

### 2.2 Automatic Code Completion

Auto-completion facilitates programming by generating predictive text that aligns with a developer's programming context.

*Granularity level*: Completion occurs at varying degrees of granularity. At the basic level, next-token prediction is applied where the model predicts the subsequent code token [1, 8, 9, 34–36]. In line completion, a model is employed to complete a line of code given a specific context [1, 3, 7, 19]. In the most extensive form of code completion, block completion, models generate entire blocks, functions, or classes [2, 3, 14].

*Completion Scenario*: Furthermore, code completion can be characterized as a left-to-right task. In this approach, the model has visibility of only the code context preceding the cursor at the point where code completion is invoked. Alternatively, it can be viewed as an editing task where the model leverages both the left and right context of a trigger point. This allows the model to generate a span of code, which can commence and conclude anywhere within a line of code, and can encompass multiple lines of code [2, 4, 34]. Our tool incorporates left-context-only models, CodeGPT and UniXcoder [3], as well as InCoder [2], a model using both left and right contexts.

### 2.3 Empirical Studies on Code Completion

*Models' performance in online settings:* Proksch et al. [37] conducted an evaluation of a method-call recommendation system using a dataset derived from IDE interactions and identified that prevalent assessments using synthetic datasets overlook changes in context. In 2019, Hellendoorn et al. [38] analyzed traditional next-token prediction models (e.g., n-grams and recurrent neural networks) using a real-world dataset from observing 66 developers' usage over the span of two weeks resulting in 15K completions. The results indicated that synthetic benchmarks lack sufficient representation. Aye et al. [18] examined neural code completion models trained on live code versus committed code. They discovered a notable decrease in model performance when evaluating the completions collected from real usage. Additionally, they observed a higher usage of out-of-vocabulary tokens in predicting completions compared to masked tokens. In a recent study, Bibaev et al. [17] improved the rankings of auto-completions in IDE's by training models on anonymized data.

*Models' performance in offline settings:* Maurasoiu et al. [39] investigated the usage of code completion in the Dart Editor by six professional software developers. Upon analyzing this data, they identified certain interaction patterns. Key findings hint that a large portion of code completions are not accepted by users, and code completion is frequently employed as a debugging tool. Ciniselli et al. [34, 35] evaluated the proficiency of two LMs, T5 [28] and RoBERTa [24], in completing code at three levels of granularity, namely single-token, line, and block completion for Java methods and Android app methods in an offline setting. Although T5 outperformed RoBERTa, the authors observed that these models' efficacy in predicting longer sequences is restricted. The outcomes of the study are specific to Java and may not be generalizable to gradually- or dynamically-typed languages. In another empirical study, Van Dam et al. [40] used the impact of contextual data on the performance of UniXcoder, CodeGPT, and InCoder at two granularity levels (token and line completion), and two languages (JavaScript and TypeScript). The results indicated that all models performed slightly better when type annotations were omitted or when multi-line comments were present. These observations suggest that care should be taken when training, fine-tuning, or selecting such models to ensure they align with the intended data and application.

*Developers' expectations of models.* Recently, Wang et al. [41] surveyed 599 programmers to understand their expectations of code completion models. They discovered that 81% of the respondents used token-level predictions to predict accurate identifier names and APIs. Additionally, 32% of participants utilized statement-level predictions, with 46% of those being related to API completions and 45% for editing the current line of code. The survey also revealed that 80% of the participants agreed that code completion tools should support identifier completions, API recommendations, and path completions. Most recently, Ciniselli et al. [42] investigated the needs of software developers in relation to code recommendation systems, surveying 80 developers and categorized 70 requirements for these systems. Notably, developers value modifiable code over creating new one, given it is time-efficient. An intuitive user interface in the recommendation system was also deemed critical. Developers demonstrated a willingness to accept longer predictions, given their high level of trust in the system's accuracy. The system's capacity to recognize not only the current project but also past ones the developer has worked on was also highlighted. Finally, the use of familiar code expressions was found to increase the system's perceived utility.

*Our study contribution.* We assess line completion models' performance, specifically left-to-right and editing models, in both *online* and *offline* settings. We enrich the limited practical assessments by methodically evaluating three recent, public transformer-based LMs for code and conducting an in-depth qualitative analysis to comprehend why developers reject completions, leading to a broad taxonomy of such reasons. Moreover, the scope of Code4Me is much larger than existing work, encompassing over 2M completions by providing service to 1,200 users for more than a year. Finally, contrary to prior research focusing on developers' expectations from optimal code recommendation systems, our study explores these models' *actual* performance in practice. Our work lays the foundation for using the failure causes we have identified in tandem with developer expectations to build stronger, more efficient completion systems.

## 3 RESEARCH QUESTIONS

In this study, our goal is to assess how well code models perform in completing lines of code, both in online and offline scenarios. We focus on line completion as it strikes a good balance between extremely short (token prediction) and excessively long completions (entire block or function completions). We have formulated the following five Research Questions (RQs) to thoroughly evaluate LMs for code both quantitatively and qualitatively.

**RQ1: Based on real-world completions, how accurate are the models?** Using Code4Me, we conduct an online evaluation to quantitatively assess the accuracy of three state-of-the-art LMs' predictions. We explore the frequency developers accept model predictions, their accuracy compared to the ground truth, and the variation in performance across different programming languages.

**RQ2: What are the most common completion trigger points, and how do the models perform on those?** Rather than being always on, auto-completion is more effective at specific trigger points [1, 38]. In RQ2, we identify the most commonly used trigger points data and assess how well models perform on them.

**RQ3: In practice, what are common scenarios in which auto-completion is invoked?** Next, we study how developers utilize auto-completion while coding, i,e., traditional left-to-right completion, editing, etc.

**RQ4: Based on real-world completions, what are the common causes of poor results?** To understand why a model underperforms, we manually examine a subset of completions and identify causes of inaccurate predictions and developer rejection.

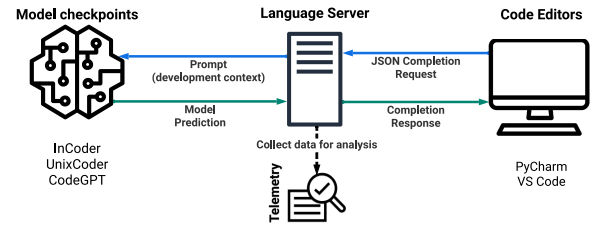

**Figure 1: Code editors and models communication pipeline**

**Figure 2: Completion sample (Code4Me PyCharm plugin)**

**RQ5: How does an online evaluation differ from offline ones?** Finally, we conduct an offline evaluation to explore how auto-completion performance varies for random and trigger masking and how these measurements compare to real-world scenarios.

## 4 CODE4ME

To gather data for answering our research questions, we developed two open-source extensions for IntelliJ and VS Code called Code4Me to test auto-completion models in an online setting. The extensions provide completions generated by three LMs and capture how developers use these completions. Code completion pop-ups are triggered upon pressing the completion key-bind or after typing one of the following trigger points [1, 22] as follow.

```
await assert raise del lambda yield return while for if
elif else global in and not or is with except . + - * /
% ** << >> & | ^ == != <= >= += -= = < > ; , [ ( { ~
```

We then request a completion using the remote API over HTTPS, which returns suggestions from all available models, and present them to developers. We utilize a standard client-server setup with an inference server running the models and a client serving as the editor for completions. Figure 1 presents the communication pipeline between the code editors, extensions, and models. Note that Code4Me competes against all completion systems (both native and non-native) in the developer's preferred code editor.

*Server.* We use a server with two GeForce GTX 2080 Ti GPUs with 22GB of memory in total to host the models. The extensions automatically make HTTPS requests to the server to receive the models' predictions. The server then feeds inputs to the models, stores relevant data, and returns the completions to the extension.

*Clients.* The clients are the local IntelliJ and VS Code extensions installed on developers' systems. Our primary requirements included (1) user-friendly operation, (2) displaying several predictions at once, and (3) integrating smoothly with the IDE without affecting its native completions. Therefore, both clients incorporate new completions into the IDE's standard suggestion window, illustrated in Figure 2. This strategy facilitates usage and enhances the existing code completions within the IDEs. Additionally, it allows for the presentation of multiple completions simultaneously. To minimize bias, we shuffle the unique completions from the three models before sending them to the client where they appear in a pop-up list, each marked with the same logo. While we ensure the presence of our completions in the list by giving them high priority in the completion provider API, it is the IDE that ultimately organizes and displays all native and non-native completions and presents them to the users. For instance, user preferences such as IntelliJ's "Sort completions based on ML", can influence the final order. During the initial phase and after collecting user feedback, we concentrated on enhancing the speed of completion generation and resolving edge cases that impacted the plugin's usability.

**IntelliJ Extension**: We built this extension by extending the existing extension template from JetBrains [43] to show custom completions in the completion window. This window shows all code suggestions to the user. The extension also detects trigger points and then prompts the models for completion. **VS Code Extension**: We developed the VS Code Extension using its API and an npm package that generates a basic extension [44]. We then extend this by adding auto-completion whenever the shortcut is used or when a trigger point is detected.

*Telemetry.* For each completion request, Code4Me collects its timestamp, trigger point, programming language, models' predictions, model inference time, user-selected prediction (empty in case of no selection), character length of left- and right-contexts, trigger kind (manual or automatic invocation), IDE, extension version, and finally the ground truth (the final line written by the developer after 30 seconds). The ground truth helps evaluate predictions based on the developer's final code use. We store the context length in characters because tokenizing for each request puts too much strain on the server/client. Additionally, users can opt-in to allow for the collection of development context (i.e., model input) to be used in RQ3 and RQ4.

*Models.* We use three recent and publicly available LMs for code: *InCoder*, *UniXcoder*, and *CodeGPT*. **InCoder** [2] is an LM for code trained on natural language and code from GitHub, GitLab, and Stack Overflow. Notably, InCoder is trained using causal masking [45], which enables it to use both left and right code contexts. **UniXcoder** [3] is an LM trained on natural language and source code using masked language modeling [23, 46], unidirectional language modeling [25], and *denoising* [28]. **CodeGPT** [19] is a GPT-2 [26] based model first introduced by Lu et al.[19], who created several GPT2-based models for code-related tasks. For our study, we create a *multilingual* version of CodeGPT by fine-tuning the pre-trained GPT-2 model checkpoint [26] on the training set of the CodeSearchNet dataset [20], which UniXcoder was also trained on [3]. This is a benchmark dataset widely used for fine-tuning models for various code-related tasks [3, 4, 9, 19, 29]. CodeSearchNet contains functions and their documentation in six programming languages: Go, Java, JavaScript, PHP, Python, and Ruby. We use a learning rate of 1.37e-4, a batch size of 2, and the AdamW optimizer with a weight decay of 0.01 for 10 epochs when fine-tuning CodeGPT. Note that we performed no further training on UniXcoder and InCoder, as these models were already trained on multilingual code. Lastly, due to latency concerns, for UniXcoder and CodeGPT, we set the beam width for decoding to 1. InCoder uses top-p nucleus sampling as opposed to beam search, hence for InCoder, we set $p = 0.95$. Additionally, to increase the speed a custom stopping condition was added to ensure models only complete the line and no extra tokens.

Table 1: Statistics on Datasets

| Language | Online #Valid Completions | Offline Random | Offline Trigger Point |
|---|---|---|---|
| Python | 219,822 | 585 | 598 |
| Java | 104,157 | 902 | 898 |
| TypeScript | 72,492 | 897 | 882 |
| PHP | 69,215 | 712 | 709 |
| JavaScript | 55,550 | 564 | 548 |
| Kotlin | 27,353 | - | - |
| C++ | 15,742 | 708 | 709 |
| Rust | 14,063 | 838 | 836 |
| C# | 12,971 | 928 | 916 |
| Go | 10,199 | 757 | 756 |
| C | 3,626 | 581 | 579 |
| Scala | 878 | 971 | 969 |
| Ruby | - | 857 | 848 |
| Total | 606,068 | 9,300 | 9,248 |

## 5 EXPERIMENTAL SETUP
### 5.1 Quantitative Analysis

*5.1.1 Models and Data.* In the offline setting, we use the Unseen dataset [13] to test the line completion performance of the three LMs. This is a curated dataset containing source code files in twelve different languages: C, C#, C++, Go, Java, JavaScript, PHP, Python, Ruby, Rust, Scala, and TypeScript. This dataset was made specifically to evaluate numerous LMs for code on data they had not been trained on. We further de-duplicate the Unseen dataset by removing repositories that are already in the training data of the models. We had previously acquired the list of repositories from the InCoder authors. This process removes 29 repositories out of the total of 293 repositories in the Unseen dataset. We use two strategies for creating the *test* sets; random and trigger masking. First, we select random positions to perform code completion, masking up to the end of the line. The selected positions are always at a whitespace character. Second, we simulate user behavior by selecting only line completions that begin at a trigger point. Both cases do not allow auto-completions at the beginning or end of a line. When generating our synthetic trigger-points, we limit the number of completions per file to 10 to prevent our results from being dominated by large files, ensuring diversity in our test samples. Table 1 displays the number of test samples per language.

*5.1.2 Evaluation Metrics.* For the quantitative analysis (RQ1, RQ2, RQ5), we use a diverse set of standard evaluation metrics [47], namely acceptance rate, ROUGE-L, BLEU4, METEOR, Edit Similarity, and Exact Match. Due to space limitations, we only report ROUGE-L here as it has been shown to align more with human judgment [48]. Additional results are available in the replication package [22]. Several of the chosen metrics necessitate the conversion of a given piece of code into a token array. For consistency in results, we utilize a uniform tokenizer, specifically the InCoder tokenizer, across all samples given its extensive exposure to numerous programming languages. The employment of the InCoder tokenizer does not confer any benefit to the InCoder model, thereby preserving the fairness of the comparison. **Acceptance Rate** is the proportion of the number of completions accepted by a user to the number of completions presented to them. **ROUGE-L** is a variant of ROUGE (Recall-Oriented Understudy for Gisting Evaluation) [49] which compares the tokenized ground truth with the prediction, evaluating it based on the longest common sub-sequence wherein each code token operates as a uni-gram. ROUGE-L calculates and employs both precision and recall to determine an F1 score. Ground truth here means the tokens that the model aims to predict. As tokens are not masked in online settings, we wait 30 seconds after offering predictions to establish ground truth. This delay allows developers time to correct any flaws in the predictions.

## 5.2 Qualitative Analysis

With RQ3 and RQ4, we aim to gain deeper insights into the strengths and weaknesses of LMs via a *qualitative* analysis of a subset of predictions generated by Code4Me in actual scenarios.

*Sampling Data.* This data is collected from Code4Me users who consented. [1] By October 2022, Code4Me had stored about 29,000 valid completions with development contexts. Due to time and cost constraints, we randomly sampled 2,117 files for manual analysis. After an initial review, we refined the completions by excluding cases where the left and right contexts were mismatched, the ground truth did not align with the context, or the model was applied to non-programming language tasks (like text documents or settings). This process yielded a total of 1,690 files for labeling.

We annotated each completion from two aspects, usage type (RQ3) and potential failure reasons (RQ4). To answer RQ3, we focus on the left and right context of an invocation without looking at the models' predictions. This RQ is aimed at evaluating whether the expectations of users (where the completion is called) align with the training settings. Next, we identify potential reasons when a completion fails. We store each completion in a file. These files include the left and right code contexts (relative to the trigger point), models' predictions, the trigger kind (manual or automatic), and the associated IDE. Additionally, we demonstrate the ground truth, representing the targeted code line after a 30-second interval, as well as the user-selected prediction. Labelers employ these files to assign fitting labels to each completion for all models.

*5.2.1 Labeling Process.* We used open coding and iteratively refined a distinct set of labels for both RQ3 and RQ4. The entire process, including defining labels, labeling samples, resolving conflicts, and creating the taxonomy, required about 540 person-hours.

For RQ3, two of the authors (referred to as $L_i$), developed a set of potential labels for invoking auto-completion. These labels represent the circumstances under which each of the three models was used. Then, two authors individually tagged each completion. Upon completion of labeling, any conflict among the labels was discussed between the labelers to achieve a consensus.

For RQ4, due to the diversity of potential failure reasons in completions, four authors undertook the process of formulating categories and assigning labels to data points. Initially, $L_1$ and $L_2$ independently examined 440 randomly selected files (20% of a model's data). Following their individual analysis, they held a meeting to form a preliminary set of potential labels together. This resulted in 22 labels. Subsequently, using this set, $L_3$ and $L_4$ engaged in individually examining another set of data points (a portion of CodeGPT predictions). The annotators analyzed available data such as left and right contexts, predictions, and ground truth. After the first iteration, annotators refined the label set to 20 labels, removing very fine-grained labels as their representation could be sufficiently covered by the remaining labels. Both $L_3$ and $L_4$ kept refining the labels iteratively until they finished the first model. Next, $L_3$ and $L_4$ had a discussion to resolve the conflicts, where several labels were merged, removed, or new ones emerged, resulting in a total of 18 labels. Following a second examination of the data, these labels were solidified and allocated to all samples. The UniXcoder and InCoder models underwent an identical iterative process, carried out by $L_3$ and $L_4$. Once again, the labels obtained from the previous model were refined throughout this procedure, to provide a comprehensive analysis of the prediction behavior demonstrated by these two models.

*Conflicts Resolution Process.* In the labeling phase, each sample was assigned an unrestricted number of labels from the annotators. To reach a consensus, annotators discussed samples that had at least one conflict and reduced the number of labels assigned to each sample to a maximum of two, selected for their ability to most accurately describe the error's origin, as determined by the experts. The process of assigning labels and resolving conflicts was conducted four times, with the labels being updated each iteration.

## 6 RESULTS

In this section, we report both the quantitative and qualitative results of this study.

### 6.1 Statistics on Code4Me

We launched Code4Me in June 2022. As of July 2023, a total of 1,203 unique users have used the extensions, with 168 of these users agreeing to share their context, thereby enabling us to conduct the qualitative analysis of the predictions. Due to our strict privacy policy that ensures user anonymity, we do not gather any personal or demographic information about our users. To limit abuse or contamination of our data we rate limit code completion to 1,000 completions per hour. The tool has responded to 1,976,701 completions, out of which 1,370,879 qualify for assessment. That is, we remove completion requests where the ground truth is empty (when the user did not write anything after calling the model), or the predictions are empty. For the online evaluation, we base our language selection on the large-scale evaluation study conducted by Xu et al. [13]. However, our users did not request enough completions for Ruby, so we replaced it with Kotlin. This selection results in twelve programming languages with 1,308,606 total data points.

---
[1] We have the approval of our institution's Ethics Review Board to conduct the qualitative analysis of the development context.

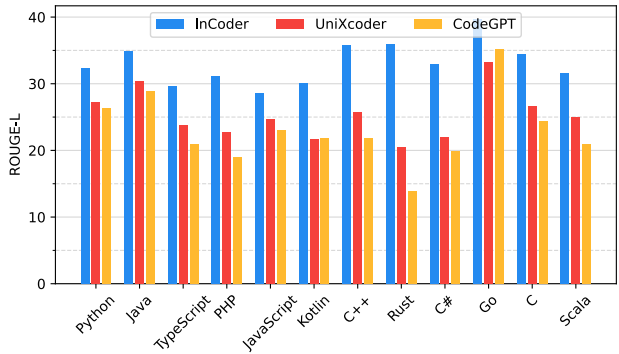

Figure 3: ROUGE-L per language for all models (online)

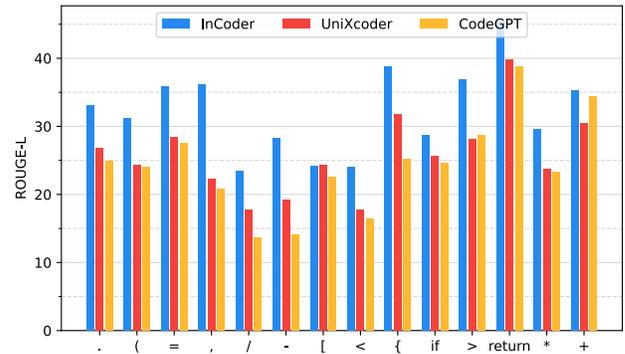

Figure 5: ROUGE-L per trigger point for all models (online)

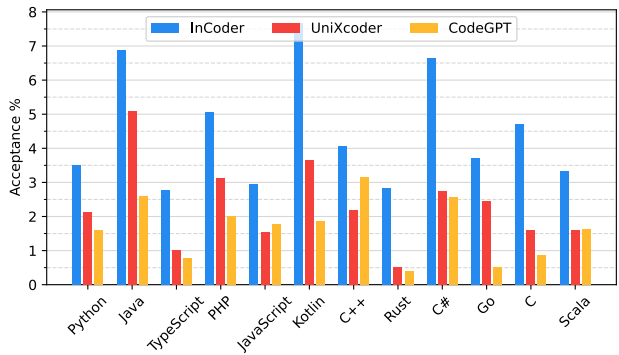

Figure 4: Acceptance rate per language (online)

Table 1 presents the number of valid data points per programming language. The IntelliJ extension generates 49.6% of the completions, while 50.4% comes from the VS Code extension.

### 6.2 RQ1: Online Evaluation

*Results on All Valid Completion Requests.* Figure 3 shows ROUGE-L of Code4Me for each language. InCoder performs best in all languages, with UniXcoder and CodeGPT closely trailing in most languages. UniXcoder and CodeGPT were trained on the same code dataset, and show similar performance. For instance, they both struggle with Rust, as it was not in their training. InCoder, however, was trained on a larger, more varied dataset that includes code understanding data such as Stack Overflow Q&A.

*Acceptance Rate.* Figure 4 presents the acceptance rate of models per language. On average, models achieve an acceptance rate of 4.91%. The lower acceptance rate of our system compared to the 10% in IntelliCode Compose can be attributed to various factors. IntelliCode Compose, during its launch, did not compete with advanced completion systems like GitHub Copilot or customized models trained on specific company or user data, which are prevalent now. In contrast, Code4Me competes with all existing completion systems within the user's code editor. Additionally, IntelliCode Compose deals with method and argument completion, which are generally simpler than line completion. It is triggered only by non-alphanumeric characters, whereas Code4Me also responds to keywords. Moreover, IntelliCode Compose, having more resources, employs a wider beam width, hence, improving the accuracy. Among models, InCoder has the highest score especially on Java, C#, and Kotlin, while CodeGPT and UniXcoder both perform exceptionally poorly on TypeScript and Rust. Note that a higher ROUGE-L score does not always correlate with increased acceptance rates. For example, Java has a higher acceptance rate than Python, despite similar ROUGE-L scores. This disparity might be due to the differing ease and verbosity of the languages. Python programmers might find it quicker to type code themselves rather than waiting for Code4Me, whereas the more verbose nature of Java makes the tool more beneficial for Java programmers. For *all valid completions*, ROUGE-L scores mostly range from 20 to 40. However, for *accepted completions*, all models score between 50 to 80 on ROUGE-L across most languages, indicating that even some accepted completions require further edits. Note that the difference between a ROUGE-L score of 50 and 80 is substantial. A score of 50 denotes moderate accuracy, suggesting correct basic structure and content but with significant gaps or errors. In contrast, a score of 80 reflects a high accuracy, with the completion closely resembling the intended output. More details are available in our public repository [22].

### 6.3 RQ2: Online Evaluation on Trigger Points

Of all completions, 83.4% come from automatic triggers set at specific trigger points. Figure 5 reports ROUGE-L scores per the 14 most frequently used trigger points for completion by Code4Me. InCoder consistently surpasses UniXcoder and CodeGPT across various trigger points. Notably, different performance is observed between these triggers. For instance, the return trigger demonstrates superior performance, potentially attributed to its positioning in a file where more contextually relevant data is available to the models. In contrast, the / trigger performs worst for all models which can be attributed to its ambiguous use in code (e.g., division, comment denotation, and file paths).

### 6.4 RQ3: Usage Scenarios

For label assignment, we analyze only the location where the prediction was triggered (left context, right context, and ground truth).

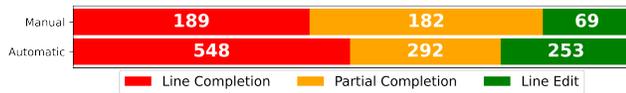

Figure 6: Auto-completion invocation: common scenarios

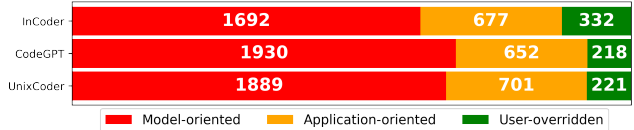

Figure 7: Top-level distribution of failure categories

This is to assess what the model would need to modify for a correct prediction. Model predictions were exclusively evaluated for RQ4. We identified three completion scenarios from the samples: (1) to predict the rest of a line based on the left context while nothing is written on the right side, (2) to partially complete a line when the model needs to predict tokens up to other tokens on the same line, and finally (3) to predict and modify the code written on the same line, in order for the correct prediction to be valid. Two annotators individually labeled all relevant files, reaching an agreement rate of 82.1%. Most disagreements arose from determining whether a line edit was necessary, which the labelers discussed to arrive at a consensus. Figure 6 presents the percentage of these scenarios. Conventional line completions stand out as the most prevalent prediction scenario (48%). However, a model exclusively trained for line completion would likely generate predictions that conflict with the right context for the remaining 52% requests. Moreover, in about 21% of situations, it would require modifications to the right context. Hence, to align with developers' needs, models should be trained to make small edits to the surrounding context.

## 6.5 RQ4: Failure Categories

After resolving all conflicts, we collected 2,800 labels for CodeGPT predictions, 2,701 labels for InCoder predictions, and 2,811 labels for UniXcoder predictions. This results in 8,312 labels for all three models. Utilizing these labels, we developed a taxonomy that encompasses all the assigned labels (Table 2). Figure 7, presents a high-level summary of label frequency per model. Our analysis indicates that completions are typically rejected due to user preference or model errors. Model errors fall into two categories: model-oriented and application-oriented. Model-oriented errors stem from the model's inherent limitations, while application-oriented errors arise from a mismatch between how models are trained to assist in development and how they are used by developers. Additionally, there are cases where the model suggests accurate or valid completions but the user prefers to override them.

### 6.5.1 Model-oriented.
The model itself is a primary source of poor predictions resulting in two main sub-categories of errors; *token-level* and *statement-level errors*.

*Token-level Errors.* Errors at the token level occur when inaccuracies arise in completing variables, functions, types, and literals.

Table 2: Taxonomy of failure categories

| Failure category plus label ID | Count |
|---|---|
| **Model-oriented Errors** | **5,511** |
| Token Level | 3,835 |
|   (ME-T1) Incorrect variable | 1,435 |
|   (ME-T2) Incorrect function | 1,162 |
|   (ME-T3) Incorrect literal | 1,130 |
|   (ME-T4) Incorrect type | 108 |
| Statement Level | 1,676 |
|   (ME-S1) Wrong parameter count | 613 |
|   (ME-S2) Wrong semantics | 352 |
|   Untimely termination | 318 |
|     (ME-S3) Early termination | 205 |
|     (ME-S4) Late termination | 113 |
|   Rambled Outputs | 249 |
|     (ME-S5) Looped repetition | 171 |
|     (ME-S6) Copied input context | 78 |
|   (ME-S7) Faulty syntax | 144 |
| **Application-oriented Errors** | **2,030** |
| (AE-1) Mid-token invocation | 1,173 |
| (AE-2) Insufficient context | 482 |
| (AE-3) Redundant invocation | 240 |
| (AE-4) Typographical errors in input | 135 |
| **User-overridden Outputs** | **771** |
| (UO-1) Correct but not accepted | 605 |
| (UO-2) Valid but not preferred | 112 |
| (UO-3) Accepted but required change | 54 |

For all models, the largest class of errors pertains to predicting variable names (ME-T1). In 28% of all files labeled with ME-T1, the model was called after the initial letters were already typed by the developer. Similarly, for 23% of all predictions labeled with ME-T2, the model was called in the middle of the identifier. Furthermore, we see a high correlation between ME-T1 and ME-T2, models that fail to correctly predict function names will also fail to predict variable names, and vice versa. Finally, in 14% of all ME-T2 labeled cases, models also predicted the wrong number of parameters (ME-S1). Incorrect literals (ME-T3) occur as frequently as ME-T2 but they mostly occur in isolation.

*Statement-level Errors.* Errors that exceed mere single-token inaccuracies and relate to a deeper comprehension of the language or code, characterized by the prediction's length, parameter count, and complex syntax.

*Number of Parameters.* A common error is the incorrect prediction of the number of parameters (ME-S1) especially in languages such as JavaScript (and dialects) and HTML/CSS as they often require a large number of parameters, often describing attributes, rather than variables. In 27% of appearances it co-occurs with an incorrect function name, in 22% of all cases with an incorrect literal, and in 20% of all cases with an incorrect variable. The parameter

```
template <typename T> T setAttribute(const key_t &key, const T
&value, access_specifier_t accessSpecifier, bool isConstant){
    if (¶
```

```
INCODER_PREDICTION: threadLock_setAttribute
UNIXCODER_PREDICTION: isConstant) {
CODEGPT_PREDICTION: isConstant)
GROUND_TRUTH: typeid(T) == typeid(key_t)) {
RIGHTCONTEXT: )
```

**Figure 8: Rejected prediction: wrong semantics (ME-S2)**

```
mine_nenerg = ['VALE3.SA', 'CSNA3.SA', 'GGBR3.SA', 'USIM3.SA']
indu_proces = ['SUZB3.SA', 'UNIP6.SA', 'SLCE3.SA', 'SMTO3.SA']
cons_civi¶
```

```
CODEGPT_PREDICTION: = ['CIVI.SA', 'CIVI.SA', 'CIVI.SA', 'CIVI.SA',
'CIVI.SA', 'CIVI.SA', 'CIVI.SA', 'CIVI.SA', 'CIVI.SA',
GROUND_TRUTH: l = ['EVEN3.SA', 'CYRE3.SA', ]
```

**Figure 9: Rejected prediction: looped repetition (ME-S5)**

```
a = int(input("enter the number to find 3 consecutive number: ")
if a % 3 == 0:
    if a % 2 =¶
```

```
INCODER_PREDICTION: 0:
UNIXCODER_PREDICTION: 0:
CODEGPT_PREDICTION: 1:
GROUND_TRUTH: = 0:
```

**Figure 10: Rejected prediction: faulty syntax (ME-S7)**

```
for i, state in enumerate(states_):
    context = state.append(previous[i])
    for c in alphebt¶
```

```
INCODER_PREDICTION: ree(self.alphabet):
UNIXCODER_PREDICTION: s(state):
CODEGPT_PREDICTION: .get_coding_handlers(context):
GROUND_TRUTH: symbols(state):
```

**Figure 11: Rejected prediction: Typographical errors in input (AE-4) and mid-token invocation (AE-1)**

values are frequently incorrect when predicting an excessive number of variables, as evidenced by ME-T1 and ME-T3 occurrences. Additionally, the number of parameters is closely related to the function name, as shown by the co-occurrence of the label ME-T2.

*Wrong Semantics.* Models make mistakes not only when completing tokens but also in the semantics of the statement they are predicting (ME-S2). Boolean expressions are often mistaken for function calls, or list comprehensions are predicted instead of indexing an array. These are errors where the model did not understand what the developer's intent was. Figure 8 shows the prediction of a variable instead of checking the equality of two types, as the condition of the `if` statement. ME-S2 is correlated with ME-T1 (26%) and ME-T2 (25%) suggesting models select identifiers based on the perceived semantics of the predicted statement.

*Untimely Termination.* LM generations can sometimes be too brief or lengthy. Early termination (ME-S3) relates to missed function calls (24%) and variables (21%), wrong literals (16%), and wrong number of parameters (13%). Similarly, models terminate late (ME-S4) due to extra variables (20%), extra function calls (19%), too many parameters (18%), and when predicting literals (16%).

*Rambled Outputs.* In these cases, the model falls into a repetitive loop of producing the same few tokens (ME-S5) until it hits the maximum allowable length, or echoes parts of the input context (ME-S6), rendering the output nonsensical. Figure 9 shows an example where the model completes the content of the array by replicating one of the strings. Potential causes are common patterns of coding or frequently-used token(s) being present in the context, specific model settings, or use of a fallback when the model is facing difficulty in predicting an appropriate response. CodeGPT exhibits fewer ME-S4 instances (28 files) compared to UniXcoder (35) and InCoder (50). Qualitative analysis shows this issue occurs more when the quality of input data is not very high. Furthermore, ME-S5 occurs more when the model copies a literal (19%) from input context.

*Faulty Syntax.* All models exhibited low incorrect syntax rates (ME-S7), with a few notable cases. ME-S7 is highly correlated (31%) with poor performance in requesting completions in the middle of tokens (AE-1), indicating potential sensitivity to out-of-distribution samples from partial identifiers. Figure 10 shows a completion request for which all models predict an assignment instead of equality in the `if` statement.

*6.5.2 Application-oriented.* Approximately 42% of completion requests are labeled with errors related how models are applied in practice and how that can lead to sub-optimal performance.

*Mid-token Invocation.* Models use tokenizers based on frequently occurring byte pairs as their smallest unit. This causes problems if a model is called in the middle of a token, as it would generate an out-of-distribution token at that location (AE-1). Our analysis shows in 34%, 23%, and 10% of cases, AE-1 co-occurs with incorrect variable names, function names, and literals, respectively. Figure 10 shows an example where models struggle to accurately disambiguate the meaning of the equals sign. This challenge is also applicable in languages such as JavaScript or TypeScript, where both == and === are employed for equality comparison, albeit with distinct functionalities.

*Insufficient Context.* Given that models use only the present file as predictive context, its content is crucial. Yet, often, context insufficiency causes errors (AE-2) in 482 predictions. For instance, early in a file, models often require importing code from specific files. This leads to a high correlation (31.3%) between AE-2 and ME-T3 labels.

*Redundant Invocation.* These are the cases where the models are unnecessarily invoked (AE-3), e.g., in a position where both correct left and right contexts already exist. We attribute this to both user behavior and implementation factors.

*Typographical Errors in Input.* Models also struggle when there are misspellings or mistakes in the left-context (AE-4). Figure 11

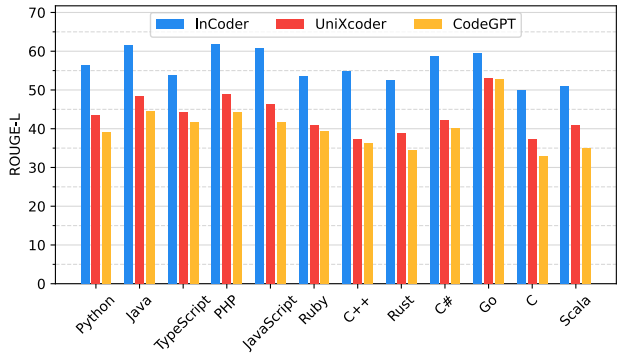

**Figure 12: ROUGE-L per language (random strategy)**

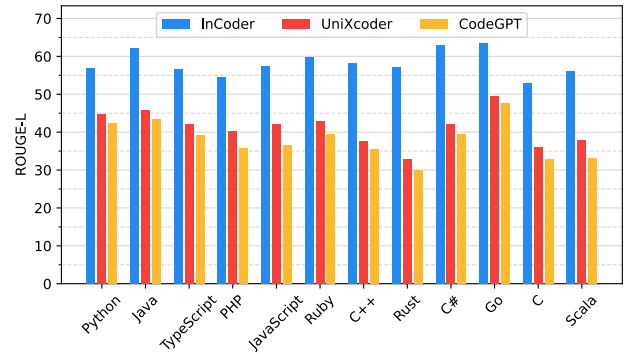

**Figure 13: ROUGE-L per language (trigger point strategy)**

presents an example where a typo in "alphebt" and a mid-token trigger cause inaccurate function prediction.

*6.5.3 User-overridden Outputs.* In these cases, users override the models' suggestions.

*Correct but not accepted.* Interestingly, the user does not always accept predictions even when they are correct (UO-1). In 605 predictions, the ground truth matched the generated predictions exactly. When examining the prediction types, we found that for 163 predictions (comprising 102 partial line completions and 61 line edits), there was context to the right of the prediction location that the UniXcoder and CodeGPT models could not have taken into account, potentially leading to overwriting. Users might also choose a completion from a competing tool, which is beyond our control.

*Valid but not Preferred.* In certain cases, although the prediction did not precisely match the ground truth, it would be a valid prediction if accepted (UO-2). The reason is likely attributed to style preferences that are currently not accounted for in the models.

*Accepted but Required Change.* In certain cases, the selected prediction does not perfectly match the ground truth (UO-3). In 22% of instances, variable modifications occurred, 13% involved literal changes, 6% witnessed function name alterations, and 9% entailed parameter count adjustments.

### 6.6 RQ5: Offline Evaluation

We report the average ROUGE-L per language for the random and trigger point test sets in Figures 12 and 13. We also report ROUGE-L per trigger point for the trigger point test set in Figure 14.

*Performance per Language.* Each model behaves differently across languages in random and trigger point tests. UniXcoder and CodeGPT, trained on identical code datasets, perform similarly in all languages. However, InCoder, trained on a larger and more diverse corpus, outperforms the rest. Its proficiency is also boosted by utilizing both left and right contexts, which can be helpful, especially in the first few lines of a file with less left but more right context.

*Impact of Masking Strategy.* Based on the results, models exhibit similar performance across both test sets. However, the main difference is that the trigger point test set yields slightly lower

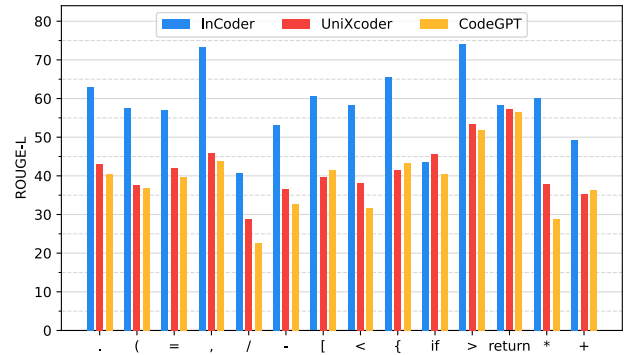

**Figure 14: ROUGE-L per trigger point**

ROUGE-L for UniXcoder and CodeGPT, whereas InCoder performs better on this test set. These differences are especially strong when considering Exact Match.

*Comparison to Online Evaluation.* Both offline test sets result in a gross overestimation of the performance of the models. While InCoder did not exceed 40 ROUGE-L for any individual language in the online evaluation, the offline evaluation shows a maximum ROUGE-L of over 60 in both the random and trigger point strategies. Metric values are similarly inflated for UniXcoder and CodeGPT, which score higher than 30 ROUGE-L in only two languages in the online setting, but achieve this for all languages in the offline setting. Though offline results are much higher than corresponding online ones, offline evaluation still holds value due to correct model rankings, i.e., InCoder followed by UniXcoder and CodeGPT. However, performance per language may not align with online settings, with slight differences in both test sets.

*Performance per Trigger Point.* Some models have proficiency gaps at certain trigger points, however, using random trigger points for auto-completion may not mimic programmers' actions well. Figure 14 shows ROUGE-L scores for three models post different trigger points. These outcomes differ greatly from our online evaluation results (Figure 5). Once again we observe that offline results are higher than the corresponding results from the online setting.

However, performance per trigger point also varies between online and offline evaluations. In online evaluation, InCoder excels at the `return` trigger point, while offline shows the three models are closely competitive. Additionally, the offline evaluation shows that the > trigger point yields very good performance while this is not the case in the online evaluation. These dissimilarities show that even a test set created based on trigger points does not perfectly match a realistic programming environment. This may be due to the fact that offline datasets often contain complete code, meaning there may be more useful code context available than in cases where a developer is actively creating and editing new code.

# 7 DISCUSSION

Below, we briefly discuss the key shortcomings of existing code completion solutions.

## 7.1 Current Limitations

*7.1.1 Models and Training Shortcomings.* Based on RQ4 results, on average for all models, around 66% of failures (excluding discarded requests), arose from model shortcomings such as inaccurate identifiers and types, semantic or syntactic errors, and untimely termination. Developers employ proper identifier names to enhance code comprehension and maintenance, however, existing models face difficulty in generating suitable names that reflect code functions and intentions due to their underlying training to predict plausible subsequent tokens, not in understanding code as developers do. Their current limitations also include inadequate knowledge of the codebase, hindering their ability to examine it, identify components, and comprehend their relationships to improve generations. Moreover, the practice of writing names and comments can vary greatly among developers, and a model trained on a diverse range of code may have difficulty learning a consistent or useful strategy for generating them. This could lead to suggestions that are confusing or unhelpful for developers.

*7.1.2 Disparity between Theory and Application.* About 24% of the failures are attributed to a misalignment between the models' practical application and their training objectives. The most prevalent instance, accounting for 58% of such cases, is the inappropriate invocation of models, either manually or automatically, at unsuitable locations like in the middle of identifiers or operators, creating new out-of-distribution tokens as input to the models. The amount and quality of context also impact these models' generations, with around 24% of instances supplying limited context, and 7% providing context with typos or syntax errors. The average left and right context length in our online data are 299 and 154 tokens. For example, relevant context might not be readily available within the file, such as functions on imported objects or a utility class with global functions. Additionally, RQ3 showed bidirectional models, need to modify the right context for the prediction to properly fit in the context. This necessity typically stemmed from brackets auto-inserted by code editors or extensive code rewrites. As anticipated, brief, poor-quality, faulty, misaligned code contexts from live code can result in substandard recommendations.

*7.1.3 Insufficient and Unrealistic Evaluation Settings.* Our quantitative analysis reveals a gap in accuracy between online and offline settings. Differences may stem from the training datasets, evaluation strategy (random token masking versus post-invocation completion), developer real-world interactions with the completion systems, individual decision-making processes, and developers' personal preferences, among other factors. For instance, the context and code quality authored by an average developer may differ from those present in the refined benchmark or proprietary datasets used in some offline evaluations. Additionally, offline data typically contain more valuable context than live development where limited context is available. In RQ5, offline evaluations did not match our online ones, making it a challenge to achieve equivalent results with both settings. Our analysis also indicates that defining successful predictions is complex. Both our research and prior studies [50] demonstrate that even accepted suggestions may not persist in a codebase as developers may initially accept a prediction and later alter it significantly. This leads to the challenge of determining the correct timing for prediction validation. Finally, RQ4 showed that in 78% of cases where developers tend to override the recommended completions (UO-1), the predictions are correct and match the ground truth exactly. Valid but not preferred completions (UO-2) are 15% of cases and in 7% of instances, the user opts for a completion that requires editing before it can be incorporated into the code (UO-3). Deciphering developers' intent from code can be tough due to its complexity and individual coding styles. Moreover, predictions may not match the ground truth but can be semantically consistent with the code and acceptable. While qualitative analysis can help, it is not a scalable solution.

## 7.2 Recommendations

Examining the outcomes and unresolved issues based on our analysis allows us to suggest several potential research directions.

*7.2.1 Align Model Training with Practice.* We propose adjusting auto-completion's training objective to decrease the granularity of masked units (e.g., characters), as well as adding partial completions, and editing capability to models. Intentionally adding bugs or small perturbations on a character level can improve performance in the presence of bugs, typos, and syntactic inconsistencies in the inputs. Next, oversampling less frequent triggers can repair the difference in performance at these points. Aligning with the expected use case can improve model performance, meet developers' preferences for intuitive auto-completion tools [42], and enhance real-time application robustness. To address the performance gap among different languages as highlighted in RQ1, RQ2, and RQ5, we recommend further research on cross-language training, identification of unique linguistic elements, and the incorporation of low-resource languages. Finally, exploring additional model architectures for this task could yield intriguing results.

*7.2.2 Reduce Model Rambling.* Overcoming rambling (ME-S5 and ME-S6) is a significant challenge in NLP. Methods like modifying the training process, using more sophisticated decoding strategies, introducing penalties during the generation process, or post-processing the generated text to remove repetitions or copied context are among the potential solutions. For instance, setting the model's *temperature* parameter to a higher value can make it less deterministic with respect to patterns from the input. Finally, using reinforcement learning, one can penalize the model for copying input directly and reward it for generating more original responses.

*7.2.3 Enrich the Context.* Relying solely on a language modeling approach for code completion presents certain limitations. For instance, it can be challenging for a model to predict user-defined types and uncommon or new APIs if they are not included in the training data (ME-T1, ME-T2, ME-T4). Additionally, if there is insufficient context within the current file, the model may struggle (AE-2). Enriching model input with additional details, e.g., available functions and imported classes could help the model better understand the context. Therefore, integrating LMs with more structured methods, such as symbolic reasoning, or incorporating data from static analyses and local context including project context (scale and domain), system context (e.g., languages and deployment environments), and personal context (e.g., expertise) can lead to more accurate modeling of user needs and higher code quality. Post-processing generations also reduce syntactic issues (ME-S7).

*7.2.4 Smart Invocation.* RQ2 and RQ4 findings highlight different issues with model invocation. To tackle issues with automatic mid-token invocations (AE-1) and redundant activation (AE-3), we can equip both models and code editors with a smart invocation capability to selectively offer suggestions. Studying developer interaction data with code editors can lead to identifying the most appropriate moments to automatically invoke auto-completion. Additionally, models should refrain from suggestions if unsure about their predictions. This will reduce noise and ensure that only relevant, high-confidence suggestions are presented to the developer.

*7.2.5 Improve Usability and Customization.* We recommend using multi-suggestion functionality (variety of viable completions) to repair valid but not accepted completions (UO-2) potentially leading to a higher acceptance rate (RQ1). While personalized recommender systems and better user interfaces can improve acceptance, further studies are needed to fully comprehend the underlying reasons for overwritten completions (UO-1, UO-2, UO-3).

*7.2.6 Improve Evaluation Settings.* The challenges of evaluating code completion systems are multifaceted including issues related to proper metrics, defining ground truth, developers' individual preferences, and their ability to alter the generated code. Strict metrics like acceptance rate can undervalue a model's worth, while fuzzy metrics (e.g., BLEU and ROUGE) overlook semantic equivalence, resulting in underestimation. Moreover, traditional offline evaluations fail to accurately assess the performance of these systems (RQ5). Online evaluations aim to address some of these shortcomings by evaluating models' performance in more realistic coding conditions. However, RQ4 showed that some *correct* completions are still rejected by developers (UO-1). Hence, an effective evaluation requires establishing a reliable ground truth and a better understanding of developers' needs. RQ3 has demonstrated that programmers employ models under varying scenarios, such as left-to-right completion or editing tasks which should be considered in the evaluation. Lastly, future work can explore the differences between the performance of IDE extensions and third-party tools.

### 7.3 Threats to the Validity

Threats to **internal validity**, such as model hyperparameters and implementation errors, could impact our study's results. We utilized public checkpoints of three LMs provided in replication packages. We have tested our implementation and chosen hyperparameters carefully, however, more optimal hyperparameters may exist. To reduce the impact of undetected errors in the code, we publish our source code and dataset to enable other researchers to replicate and extend our work. To minimize subjectivity bias, four authors participated in open coding. In every round, at least two authors independently annotated samples using available code context. Any arising conflicts were subsequently resolved through discussion. These processes increase our confidence in the taxonomy, however, as subjectivity is unavoidable in qualitative analysis, further large-scale studies can refine or verify our taxonomy.

Threats to **external validity** include dataset quality and result generalizability. For enhancing generalizability, we utilize three public code LMs with different training objectives and architectures, and 12 programming languages based on the literature [13]. This allows representation across a diverse set of models and languages. In offline evaluation, we utilized and deduplicated a standard dataset [13] to prevent data leakage from the training data [51]. For fine-tuning CodeGPT on multilingual data, we used another common benchmark dataset [20]. A potential concern is that the Code4Me user base may not reflect the broader programming community. To address this, we have ensured extensive distribution of our tool to expand our user base over a year. We believe it mitigates the potential bias, resulting in a representative dataset. Yet, further studies can verify and generalize our findings to more data, models, languages, and users.

**Construct validity** threats relate to the appropriateness of our metrics. To evaluate Code4Me, we use a diverse set of widely-used metrics from the literature [1, 7, 34, 35, 50, 52], however, due to space constraints, we only report ROUGE-L and acceptance rates here, with the rest available in our replication package [22].

## 8 CONCLUSION

In this paper, we evaluate the use of Transformer-based models for code completion in a practical setting. To this end, we developed Code4Me, an open-source completion extension that can work with three popular LMs. Over the course of a year, we collected over 600K valid completions from 1,200 developers. Our empirical analysis demonstrates that language popularity, trigger points, and completion scenarios substantially affect accuracy. Moreover, the prevalent use of offline evaluations in existing research tends to paint an overly optimistic picture of the current auto-completion capabilities. Our findings suggest a need for a focused study on auto-completion that considers real-world applications, as well as training LMs that better align with the needs of developers.

### ACKNOWLEDGMENTS

Our special thanks go to Frank van der Heijden and Jorit de Weerdt whose support and assistance played an important role in Code4Me's success. We also thank the users who used Code4Me in their programming activities. Finally, we thank Dr. Georgios Gousios for providing hardware resources for part of the experiments.